\documentclass[fleqn,12pt,twoside]{article}
\usepackage[headings]{espcrc1}

\readRCS
$Id: espcrc1.tex,v 1.2 2004/02/24 11:22:11 spepping Exp $
\usepackage{graphicx}
\usepackage[figuresright]{rotating}

\newcommand{\ds}[1]{#1 \hspace{-0.5em}/}  
\newcommand\bzeta{\mbox{\boldmath$\zeta$}}

\newcommand{\AmS}{{\protect\the\textfont2
  A\kern-.1667em\lower.5ex\hbox{M}\kern-.125emS}}

\title{Ferromagnetism in quark matter and origin of the magnetic field in
compact stars}

\author{T. Tatsumi\address[KU]{Department of physics, 
        Kyoto University, Kyoto 606-8502, Japan},%
        T. Maruyama\address{College of Bioresource Sciences, Nihon
        University, Fujisawa 252-8510, Japan.},
        E. Nakano\address{Yukawa Institute for Theoretical Physics,
        Kyoto University, Kyoto 606-8502, Japan}
        and
        K. Nawa\addressmark[KU]}

\begin{document}

\maketitle

\begin{abstract}
Two magnetic aspects of quark matter, ferromagnetism and spin density
 wave, are discussed in the temperature-density plane. 
Some implications of ferromagnetism are suggested on relativistic 
heavy-ion collisions and compact stars.
\end{abstract}

\section{INTRODUCTION}

So far, many studies have been devoted to understand high-density and/or
high-temperature QCD. In this talk we are concentrated on magnetic
aspects of quark matter. We shall discuss the possibility of 
ferromagnetism and spin density wave in quark
matter. These magnetic properties of
quark matter should be closely related to phenomena observed in
relativistic heavy-ion collisions or compact stars. Actually our study
has been stimulated by the recent discovery of magnetars \cite{tho}, which have
huge magnetic field of $10^{15}$G and come into a new class of compact stars.  
It has been first estimated by $P-\dot P$ curve for pulsars and more recently
suggested by observation of the cyclotron absorption lines.  

The origin of the magnetic field in compact stars has been a
long-standing problem since the first discovery of pulsars in early
seventies. Instead of asking for the origin, many people have taken a
simple working hypothesis, conservation of magnetic flux during the
stellar evolution. However, if we apply this hypothesis to explain the
magnetic field of magnetars, we immediately have a contradiction that their
radius should be much less than the Scwartzschild radius. Thus, the
discovery of magnetars seems to give a chance to reconsider the origin
of the magnetic field in compact stars. Since there is widely developed
hadronic matter inside compact stars, we 
consider the spin-polarization in hadronic matter as 
a microscopic origin of the magnetic field in compact stars.
 
\section{RELATIVISTIC FERROMAGNETISM}

We first studied the possibility of ferromagnetism in quark matter in 
the perturbative way, in analogy with the Bloch mechanism for itinerant
electron system \cite{tat}. Consider the spin-polarized electron gas 
interacting with the Coulomb force. 
Then total kinetic energy is
increased, while the interaction energy is decreased due to the Pauli
principle; an electron pair with the same spin cannot approach each
other to effectively avoid the Coulomb repulsion. 
\vskip -0.8cm
\begin{figure}[h]
\begin{minipage}{0.51\linewidth}
~~~In the relativistic theories, spin is not a good quantum number. 
Still, we can consider a ferromagnetic state
with recourse to the Pauli-Luvansky vector and the spin four vector
$a^\mu$
, which is specified not only by the spin direction
vector $\bzeta$ in
the rest frame of each quark but also by its momentum $\bf p$, different
 from non-relativistic theories. 
Thereby there are many choices for
the spin configuration in the phase space of quark matter,
which are all reduced to a constant $\bzeta$ configuration in 
the non-relativistic limit \cite{mar}.
Using the standard spin configuration given by the Lorentz
transformation of $\bzeta=\pm {\hat{\bf z}}$ for each quark, we can 
see a weakly first-order phase
transition to ferromagnetic state at low densities $n_q$ 
 (Fig.~1). We also see that there is metamagnetic state even in the
 paramagnetic phase.
\end{minipage}
\hspace{\fill}
\begin{minipage}{0.46\linewidth}
\begin{center}
\includegraphics[width=6.5cm]{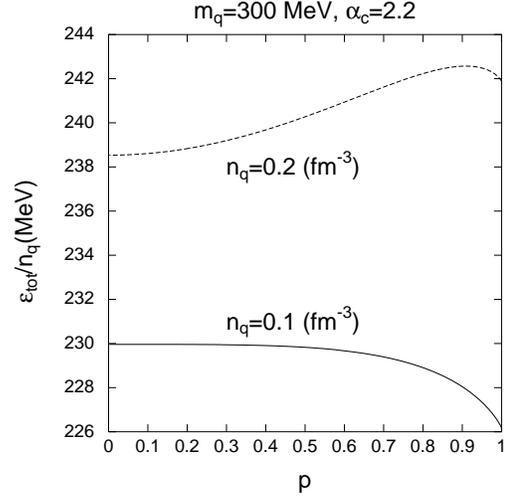}
\end{center}
\vskip -1cm
\caption{First-order phase transition with MIT bag-model
 parameters. $p\equiv (n_+-n_-)/n_q$}
\end{minipage}
\end{figure}
\vskip -0.8cm
To get more insight into the ferromagnetic transition we must treat it in a
non-perturbative way; the Hartree-Fock approximation may be the next
step \cite{naka,nak}. When we consider the Fock self energy $V$ for quarks, 
the propagator
can be simply written as
\begin{equation}
G^{-1}(p)=\ds{p}-m_q+V(p),
\end{equation}
where the self-energy consists of the mean-fields in various
channels. Among them the axial-vector mean-field ${\bf V}_A$ is responsible to
magnetization. Taking only ${\bf V}_A$ for simplicity along the $z$-axis,
$V_A=-\gamma^5\gamma^3U_A$, we find the single-particle energy,
\begin{equation}
\epsilon_n=\pm\left[{\bf p}^2+m_q^2+U_A^2+\zeta U_A\sqrt{m_q^2+p_z^2}\right]^{1/2},
\label{energy}
\end{equation}
for each spin state $\zeta=\pm 1$. Here we can clearly see that it 
is split depending on the spin state 
\footnote{It is called as the exchange splitting in the non-relativistic case.}
and rotation
symmetry is broken due to the coupling of momentum and ${\bf V}_A$. Thus
we have two Fermi seas with different shapes: the Fermi sea of majority
particle is deformed in the ``prolate'' shape, while that of minority
particle in the ``oblate'' shape \cite{naka}. Since $U_A$ should be given by $G(p)$,
\begin{equation}
U_A(k)=-\frac{g^2}{2}\frac{N_c^2-1}{4N_c^2}\frac{1}{N_f}\int\frac{d^4p}{(2\pi)^4}D(k-p){\rm tr}\left[\gamma^5\gamma^3G(p)\right],
\label{ua}
\end{equation} 
it may still be a complicated task to solve Eq.~(\ref{ua}) in a
self-consistent manner. Here we use
an effective model with a {\it zero-range} interaction instead of the original OGE,
\begin{equation}
{\cal H}_{\rm int}=G^2{\bar q}\gamma_\mu\gamma^5q{\bar q}\gamma^\mu\gamma^5q.
\end{equation}
Evaluating Eq.~(\ref{ua}) over not only the Fermi seas but also the Dirac sea, 
we find that the phase transition is of second order at the critical chemical potential,
\begin{equation}
\mu_c=\mu_c^F\Lambda/m_q e^{-\gamma_E/2},~~~\mu_c^F={\rm exp}\left[\frac{4\pi^2}{3N_fN_cG^2m_q^2}\right],
\label{crit}
\end{equation}
for $m_q\ll\Lambda$, where $\mu_c^F$ is given by the genuine Fermi-sea effect and other
factors by the vacuum polarization effect after the proper-time regularization
with a cutoff parameter $\Lambda$\cite{nak}. We can clearly see the effect of quark mass as
well as a non-perturbative nature
with respect to the coupling strength. 
Thus quark matter is in a ferromagnetic phase above $\mu_c$.
It would be interesting to compare this result with the previous one
given by the perturbative calculation, where we have seen that
a ferromagnetic phase develops at rather low densities. This difference
should come by the difference of the interaction range: the infinite
range in the latter case and the zero range in the present case. The
order of the phase transition is also changed from the first order to
the second order. These qualitative differences may be very important and
need further investigations \cite{maru}.

\section{COLOR MAGNETIC SUPERCONDUCTIVITY}

When the ferromagnetic phase is realized in quark matter, its relation
with the color superconductivity should be elucidated. We have discussed
the possibility of the coexistence of both phase in quark
matter \cite{naka}. Starting with the OGE interaction again, we leave the
particle-particle pair mean-field $\Delta$ as well as the particle-hole
one $V$; the former
is responsible to color superconductivity, while the latter
ferromagnetism. Considering the $\bar 3$ pair field both in the color and flavor
space for $\Delta$, we assume quark pairs with the same spin state
$\zeta=\pm 1$ in each Fermi surface. Then we found that the gap functions
$\Delta_\zeta$ have angular dependence on the Fermi surfaces like in the $A$
phase of the liquid $^3$He: they take nodes
on the both poles and maximum around the equator. Consequently we have
shown that the coexistence of ferromagnetism and color superconductivity
is possible, while the latter somewhat suppresses magnetism \cite{naka}.

\section{CHIRAL SYMMETRY RESTORATION AND SPIN DENSITY WAVE}

In the recent papers we have discussed another magnetic aspect in the
intermediate densities \cite{nakan}, where chiral symmetry restoration 
has been
expected. We consider a density-wave state with
\begin{equation}
\langle {\bar q}q\rangle=\Delta{\rm cos}{\bf q}\cdot{\bf r},~~~
\langle {\bar q}i\gamma^5\tau_3 q\rangle=\Delta{\rm sin}{\bf q}\cdot{\bf r},
\label{dcdw}
\end{equation}
for two-flavor quark matter. 
Then we can see that the non-trivial spatial 
dependence of the chiral angle $\theta={\bf q}\cdot{\bf r}$ induces the
kinetic term of the density wave $\propto q^2$ by the vacuum polarization as well as the
axial-vector coupling term with quarks. The latter term gives an energy
gain due to the {\it nesting} of the Fermi surface
If it is superior to the energy increase due to the former effect, such a
density-wave state is favored. We call it as {\it dual chiral-density wave}
(DCDW) state. Physically the amplitude gives a dynamical mass to
quarks and is closely related to the
chiral-symmetry restoration, while the phase gives a magnetic ordering
to quark matter; the mean spin value vanishes in this phase, 
but  total magnetic moment is spatially oscillating. So, this phase is a
kind of the {\it spin density wave}. 
We have explored the possible region of the DCDW state in
the temperature-density plane, by explicitly using the NJL model
\cite{nakan}. It
should be also interesting to consider that the ansatz (\ref{dcdw}) describes another
path for chiral-symmetry restoration, different from the usual one 
(${\bf q}=0$). In this view chiral-symmetry restoration is delayed by
the presence of DCDW.

\section{SUMMARY AND CONCLUDING REMARKS}

We have discussed two magnetic aspects expected in quark matter:
ferromagnetism and spin density wave. For the latter we may notice a
resemblance of DCDW to pion condensation in hadron matter. Actually, the
symmetry structure is common, and thereby magnetic property is the same: 
the pion condensed phase has been known to have anti-ferromagnetic order \cite{pio}.
So, one may expect a hadron-quark continuity between two phases.

Ferromagnetism has some implications on phenomena in relativistic
heavy-ion collisions, cosmology or compact stars. If the strange matter
is absolutely stable, strangelets should be magnetized like small
magnets, or 
magnetized quark lumps may be left as relics of QCD phase
transition in early universe to produce a primordial magnetic field in
the galaxies. More direct implication may be given for magnetic fields in
compact stars. Assuming that magnetars are mostly occupied by quark matter,
we can easily estimate the magnitude of the surface magnetic field $B$ to be 
$O(10^{15-17}{\rm G})$\cite{tat}. As another implication, ferromagnetism may give a
scenario about the hierarchy of the magnetic field observed in compact
stars;  there are located three colonies of pulsars with different $B$ 
in the $P-\dot P$ plane.
One may need some dynamical effects to explain it like
destruction 
and formation of magnetic domain or metamagnetic state, depending on
change of the
external parameters such as rotation, mass accretion or temperature.

This work is supported by the Japanese Grant-in-Aid for Scientific
Research Fund of the Ministry of Education, Culture, Sports, Science and
Technology (13640282,16540246). It is also partially supported by the
Grant-in-Aid for the 21th Century COE ``Center for the Diversity and
Universality in Physics'' from the Ministry of Education, Culture,
Sports, Science and
Technology of Japan.

\end{document}